\documentclass[manuscript]{emulateapj}

\usepackage{graphicx}
\usepackage{dcolumn}
\usepackage{bm}
\usepackage{xcolor}
\usepackage{color}
\usepackage{verbatim}
\usepackage{lineno}
\usepackage{amsmath}
\usepackage[hang]{footmisc}

\usepackage{floatrow}
\usepackage[title]{appendix}

\newcommand{\mcmc}{Dust2Dust}
\newcommand{\mass}{$M_{\rm stellar}$}

\newcommand{\logl}{$\mathcal{L}$}
\newcommand{\mures}{$\mu_{\rm res}$}

\newcommand{\burnin}{100}
\newcommand{\numsim}{400}
\newcommand{\EBV}{$E(B - V)$}

\newcommand{\sigint}{$\sigma_{\rm int}$}

\newcommand{\RMS}{$\sigma_{r}$}
\newcommand{\chilowres}{\chi^{2}_{\mu_{\rm res}, \rm low}}
\newcommand{\chihires}{\chi^{2}_{\mu_{\rm res}, \rm high}}
\newcommand{\chilowrms}{\chi^{2}_{\sigma_{\rm r}, \rm low}}
\newcommand{\chihirms}{\chi^{2}_{\sigma_{\rm r}, \rm high}}

\newcommand{\covbinw}{0.011}

\def\MELHcm{$-0.077 \pm 0.006$}
\def\MELHcs{$0.058 \pm 0.005$}
\def\MELLRm{$3.026 \pm 0.375$}
\def\MELLRs{$1.481 \pm 0.418$}
\def\MELHRm{$2.138 \pm 0.25$}
\def\MELHRs{$1.061 \pm 0.429$}
\def\MELLTe{$0.087 \pm 0.016$}
\def\MELHTe{$0.107 \pm 0.018$}
\def\MEHLTe{$0.086 \pm 0.021$}
\def\MEHHTe{$0.114 \pm 0.015$}
\def\MELHbm{$2.064 \pm 0.174$}
\def\MELHbs{$0.308 \pm 0.08$}

\def\cmean{$\overline{c}$}
\def\cstd{$\sigma_{c}$}
\def\betamean{$\overline{\beta}_{\rm SN}$}
\def\betastd{$\sigma_{\beta_{\rm SN}}$}
\def\RVmean{$\overline{R}_V$}
\def\RVstd{$\sigma_{R_V}$}
\def\tautau{$\tau_{E}$}

\begin{document}

\preprint{APS/123-QED}

\title{The Pantheon+ Analysis: Forward-Modeling the Dust and Intrinsic Colour Distributions of Type Ia Supernovae, and Quantifying their Impact on Cosmological Inferences}
\author{Brodie Popovic\footnotemark[1], Dillon Brout\footnotemark[2,3], Richard Kessler\footnotemark[4], Daniel Scolnic\footnotemark[1]}
\affiliation{$^1$Department of Physics, Duke University, Durham, NC, 27708, USA.}
\affiliation{$^2$ Center for Astrophysics, Harvard \& Smithsonian, 60 Garden Street, Cambridge, MA 02138, USA}
\affiliation{$^3$ NASA Einstein Fellow}
\affiliation{$^4$Department of Astronomy and Astrophysics, The University of Chicago, Chicago, IL 60637, USA.} \affiliation{$^5$Kavli Institute for Cosmological Physics, University of Chicago, Chicago, IL 60637, USA.}

\begin{abstract}

Recent studies have shown that the observed colour distributions of Type Ia SNe (SNIa) can be well-described by a combination of a dust distribution and an intrinsic colour distribution. Using the Pantheon+ sample of 1701 SNIa, we apply a new forward-modeling fitting method (\mcmc) to measure the parent dust and colour distributions, including their dependence on host-galaxy mass. At each fit step, the SNIa selection efficiency is determined from a large simulated sample that is re-weighted to reflect the proposed distributions.  We use five separate metrics to describe the goodness-of-fit: distribution of fitted light-curve colour $c$, cosmological residual trends with $c$, cosmological residual scatter with $c$, fitted colour-luminosity relationship $\beta_{\rm SALT2}$, and intrinsic scatter \sigint. We present the results and the uncertainty in 12-dimensional space. Furthermore, we measure that the uncertainty on this modeling propagates to an upper threshold uncertainty in the equation-of-state of dark energy $w$ of 0.014(1) for the Pantheon+ cosmology analysis, and contributes negligible uncertainty to the Hubble constant H$_0$. The \mcmc\ code is made publicly available.
\end{abstract}

\maketitle

\section{Introduction}

\footnotetext[1]{Email: brodie.popovic@duke.edu}

The accelerating expansion of the universe was discovered by \cite{Riess98} and \cite{Perlmutter99} using Type Ia Supernovae (SNIa) to measure cosmic distances. `Dark Energy', which is commonly parameterised by an equation-of-state $w$, is a possible explanation for this expansion and still an outstanding cosmological mystery today. To standardise the SNIa brightness for accurate distance measurements, some of the early analyses attempted to separate the components of the measured colour between intrinsic and Milky-Way like dust extinction (e.g., \citealp{Riess96} MCLS). However, over the last decade, the majority of cosmological analyses with SNIa (e.g. \citealp{Conley11,Betoule14,Scolnic18, Jones18, Brout19b}) and measurements of scatter (\citealp{Guy10}, G10; \citealp{Chotard11}, C11) have used the SALT2 approach that is agnostic to different components of colour and treats it as a single parameter \citep{Guy10}.

While larger, more modern samples with improved calibration have led to the reduction of many systematic uncertainties in cosmological analyses with SNIa, there remain lingering issues about how to best standardise SNIa. These include what has colloquially been named `the mass step'; that standardised SNIa are typically $\sim 0.05$ magnitudes brighter in heavier galaxies $( > 10^{10} M_{\odot}$ ) than their lighter counterparts \citep{Sullivan10, Kelly10, Lampeitl10}.  Furthermore, it is not understood why the empirically measured reddening law for SNIa differs from that measured for the Milky Way, particularly in the ultraviolet wavelength region \citep{Betoule14, Amanullah15}. Recently, \cite{BS20} (hereafter BS21), building off of earlier works investigating the effects of dust on SNIa \citep{Folatelli10, Chotard11, Burns14, Mandel17}, proposed introducing dust extinction to the SNIa SALT2 model to include both intrinsic colour variation within SNIa as well as dust effects tied to host-galaxy properties. This dust-based approach resulted in simulations that predict the mass step due to a difference in effective colour-luminosity relationships based on host-galaxy mass \citep{Popovic21a}, and also predict the previously unexplained increase in SNIa Hubble scatter for redder events \citep{BS20}.

While a simulation with the BS21 model can replicate the mass step in the optical bands, other studies on the effects of dust on SNIa have provided different results. One prediction of the BS21 model is that a mass step in the rest-frame Near-InfraRed (NIR) should be significantly smaller than in the optical because the NIR is less sensitive to dust.  Studies of the mass step in the NIR have been inconclusive: \cite{Ponder20} and \cite{Uddin20} find evidence of a post-standardisation mass step in the NIR, despite the negligible effects of dust in the NIR; while \cite{Johansson20} show a post-standardisation mass step in the optical band but not the NIR. Additionally, \cite{Johansson20} find evidence of a difference in dust  distributions between high and low mass galaxies, though this difference is smaller than that found in BS21.

However, \cite{BS20} left several key issues unaddressed. The original distribution of host-galaxy stellar mass assumed equal numbers of high and low mass galaxies, leading to biases in recovered parameters. Moreover, the model parameters for BS21 were determined via a coarse grid $\chi^2$ search, and included neither robust uncertainty measurements nor covariance between fitted parameters. Finally, the SALT2+dust model has not been re-trained to adjust the intrinsic SNIa colour law, nor the SED templates. While BS21 showed significant model improvements, these unaddressed issues, alongside the lack of an effective bias-correcting methodology, are significant impediments to the adoption of BS21 as a scatter model in cosmological analyses.

This paper aims to address the first two issues by describing a likelihood approach to determine model parameters, robust uncertainty calculations, and covariances between parameters. This is accomplished with the use of robust and realistic simulations of supernova surveys generated by the SuperNova Analysis (SNANA) program \citep{SNANA} in a novel combination with an importance sampling Markov-Chain Monte Carlo method: \mcmc. This method builds on approaches such as those in \cite{Scolnic16}, \cite{Popovic21a} to forward model SNIa parameters and improves the likelihood. The ideal approach to inferring model parameters is to create new simulations with updated parameters for each iteration. This is both computationally inefficient and time intensive. \cite{Scolnic16} and later \cite{Popovic21a} eluded this computationally intensive problem by creating simulations with uniform distributions of SNIa parameters, and creating a `migration matrix' to infer parent populations of stretch and colour; here we further elaborate on this method with a fast forward-modeling methodology (Section \ref{sec:fastforward}). Model retraining of the intrinsic colour law and Spectral Energy Distribution templates will be developed in a later work.

Section 2 presents an overview of the SALT2 and BS21 models. In Section 3, we provide an overview of the data and simulations. The methodology of \mcmc\ is described in Section 4. Results and the impact on cosmological measurements detailed in Section 5. Finally, we provide a discussion in Section 6 and acknowledgements in Section 7.

\section{Model Overview}\label{sec:Model}

SNIa are analysed with the use of a light-curve fitting program. Here we review the SALT2 framework and the dust components from BS21.

\subsection{SALT2}\label{sec:Model:subsec:SALT}

We use the SALT2 model as presented in \cite{Guy10} and the trained SALT2 model parameters from \cite{Taylor21}. The SNIa flux is given by SALT2 as 
\begin{align} 
\label{saltmodel}
\begin{split}
F(\rm{SN}, p, \lambda) = x_{0} &\times\left[M_{0}(p, \lambda)+x_{1} M_{1}(p, \lambda)+\ldots\right] \\
&\times \exp [c C L(\lambda)],
\end{split}
\end{align}
where $x_0$ is the overall amplitude of the light-curve, $x_1$ is the observed light-curve stretch, and $c$ is a parameter describing the colour of the SNIa. The $M_0$, $M_1$, and $CL(\lambda)$ parameters are global model parameters determined from a training program \citep{Guy10} that uses a large set of photometry and spectra; $M_0$ is the average Spectral Energy Distribution (SED) at each phase, $M_1$ describes the $x_1$-dependence of SED variability, and $CL(\lambda)$ is the average colour law. 

We infer distances following the Tripp estimator \citep{Tripp98}, the distance modulus $\mu$ is found by
\begin{equation}
\label{eq:tripp}
    \mu = m_B + \alpha_{\rm SALT2} x_1 - \beta_{\rm SALT2} c - M(z_i)
\end{equation}
where $m_B = -2.5\textrm{log}_{10}(x_0)$; $x_1$ and $c$ are defined above, and $\alpha_{\rm SALT2}$ and $\beta_{\rm SALT2}$ are global nuisance parameters for the stretch-luminosity and colour-luminosity relationships respectively, following \cite{Guy10}. $M(z_i)$ is the absolute magnitude of a SNIa with $c=x_1=0$ from a \citep{Marriner11} fit. We do not include an ad-hoc step based on mass, as \cite{Popovic21a} finds that a difference in $R_V$ distributions between high and low mass galaxies recreates the observed luminosity difference. 

\subsection{BS21}\label{sec:Model:subsec:BS21}
Here we present a review of the BS21 model. BS21 attributes observed SNIa colours to two components: a colour component intrinsic to SNIa properties, $c_{\rm int}$, and a dust component described by a distribution of reddening values drawn from the extinction ratio $R_V$ ($E_{\rm dust}$). The observed colour, $c_{\rm obs}$, is modeled as 
\begin{equation}
    c_{\rm obs} = c_{\rm int} + E_{\rm dust} + \epsilon_{\rm noise}.
\label{eq:cobs}
\end{equation}
where $\epsilon_{\rm noise}$ is measurement noise otherwise unaccounted for \citep{BS20}. This approach leaves arbitrary choices for parametric modeling of distributions. Following BS21 we use the following 7 parameters for SNIa:
\begin{itemize}
    \item \cmean: the Gaussian mean of intrinsic colour distribution.
    \item \cstd: the Gaussian sigma of intrinsic colour distribution.
    \item \betamean: the Gaussian mean of distribution for intrinsic colour-luminosity correlation.
    \item \betastd: the Gaussian sigma of distribution for intrinsic colour-luminosity correlation.
    \item \RVmean: the Gaussian mean of $R_V$ distribution.
    \item  \RVstd: the Gaussian sigma of $R_V$ distribution.
    \item \tautau: describes the exponential distribution for $E_{\rm dust}$.
\end{itemize}   

Note that $\beta_{\rm SN}$ and $\beta_{\rm SALT2}$ are different parameters.  $\beta_{\rm SALT2}$, along with $\alpha_{\rm SALT2}$, are determined from a global fit of the fitted SALT2 parameters. In the BS21 model, $\beta_{\rm SALT2}$ is a convolution of $\beta_{\rm SN}$ and other dust effects. Similarly, the observed colour distribution, $c_{\rm obs}$, is described by a symmetric intrinsic distribution combined with a dust model that accounts for the red tail observed in the data. The $c_{\rm obs}$ distribution can be phenomenologically replicated with an asymmetric Gaussian as done in \cite{Scolnic16} or \cite{Popovic21a}. 

The dust reddening component $E_{\rm dust}$ from Eq.~\ref{eq:cobs} is interpreted as \EBV\ so that the $V-$band extinction is given by 
\begin{equation}
    A_V = R_V \times E_{\rm dust}
\label{eq:avdustmodel}
\end{equation}
where $R_V$ is selective extinction and $E_{\rm dust} = E(B-V)$. The change in observed brightness is modeled as

\begin{equation}
    \Delta m_B = \beta_{\rm SN}c_{\rm int} + (R_V+1)E_{\rm dust} + \epsilon_{\rm noise}.
    \label{eq:deltamb}
\end{equation}

Following \cite{Riess96} and \cite{Jha07}, these $E_{\rm dust}$ values are drawn from an exponential distribution with probability 
\begin{equation}
    P(E_{\rm dust}) = 
    \begin{cases} 
      ~\tau_{E}^{-1}e^{-E_{\rm dust}/\tau_E} & ,~E_{\rm dust} > 0\\
      ~0 & ,~E_{\rm dust} \leq 0
   \end{cases}
\label{eq:dustprob}
\end{equation}
where $\tau_E$ is described above.

\begin{table}[!t]
\scalebox{1}{%
\begin{tabular}{c|c}
    BS21 Param & $N_{\rm par}$  \\
    \hline
    Gaussian $c_{\rm int}$ & 2 \\
    Gaussian $\beta_{\rm SN}$ & 2 \\
    Gaussian $R_V$ (low mass) & 2 \\
    Gaussian $R_V$ (high mass) & 2 \\
    Exponential $E_{\rm dust}$ (low-$z$, low mass) & 1 \\ 
    Exponential $E_{\rm dust}$ (low-$z$, high mass) & 1 \\ 
    Exponential $E_{\rm dust}$ (high-$z$, low mass) & 1 \\ 
    Exponential $E_{\rm dust}$ (high-$z$, low mass) & 1 \\ 
    \hline
    Total & 12 
\end{tabular}%
}
\caption{Fitted Parameters in BS21 Model}
\label{tab:Params}
\end{table}

Seven unique parameters are required to describe the BS21 scatter model. To account for host-galaxy correlations, we split the $R_V$ and $E_{\rm dust}$ distributions on host-galaxy stellar mass, specifically across galaxies with \mass $> 10^{10} M_{\odot}$ (high mass) and those with \mass $< 10^{10} M_{\odot}$ (low mass). Following BS21, the $E_{\rm dust}$ distributions were split between low $z$ surveys and high $z$ surveys as well as mass; however, the $c_{\rm int}$ and $\beta_{\rm SN}$ distributions are not split on host-galaxy stellar mass, as the model predicts that $c_{\rm int}$ and $\beta_{\rm SN}$ are intrinsic to SNIa and not their environments.

This splitting raises the number of fitted model parameters to 12. An accounting of the parameters and summary of the number of dimensions is shown in Table \ref{tab:Params}.

\begin{table*}[!t]
\scalebox{1}{%
\begin{tabular}{c|c|c|c}
    Survey & Cadence & DETEFF & SPECEFF  \\
    \hline
    SDSS & \cite{Kessler13} & \cite{SNANA} & \cite{Popovic21a} \\
    DES & \cite{Smith20}  & \cite{Kessler19} & \cite{DES3YR} \\
    PS1 & \cite{Jones18}  & \cite{Jones17} & \cite{Scolnic18} \\
    SNLS & \cite{Kessler13}  & N/A & \cite{Popovic21a} \\
    HST & \cite{Scolnic18} & N/A & N/A  \\
    Foundation & \cite{Jones18}  & N/A & \cite{Jones18} \\
    Low-z & Derived from Data\footnotetext[1]{See \cite{Kessler19}}$^{\rm a}$ & \cite{Kessler19} & \cite{Scolnic18} \\
\end{tabular}%
}
\caption{Source of Instrumental Inputs to SNANA Simulation}
\label{tab:SimInputTable}
\end{table*}

\section{Data, Simulation, and Selection}\label{sec:data}

Here we provide an overview of the data, simulations, and selection requirements. While simulations are typically used for bias corrections (such as BEAMS with Bias Corrections, \citealp{Kessler16}) in cosmological analyses, here we leverage the ability of simulations to provide large samples with known truth values in order to forward-model the BS21 parameters in our data.

\subsection{Data}

For this analysis, we use the upcoming Pantheon+ sample (Scolnic et al. in prep), a collection of publically available and spectroscopically classified photometric light-curves of SNIa. The low-redshift sample is compiled from the Center for Astrophysics SNIa data sets (CfA1-4, \cite{cfa1,cfa2,Hicken09b,Hicken09a,Hicken12}), the Carnegie Supernova Project (CSP,  \cite{Stritzinger2010, Krisciunas17}), the Swift Optical Archive (SWIFT, \cite{swift}), Foundation \citep{Foley17}, the Lick Observatory Supernova Search (LOSS, \cite{Stahl19,kaitmo}), and the Complete Nearby Low Redshift Supernova Sample (CNIa0.02, \cite{CNIa}).

\begin{figure}[h]
\includegraphics[width=8cm]{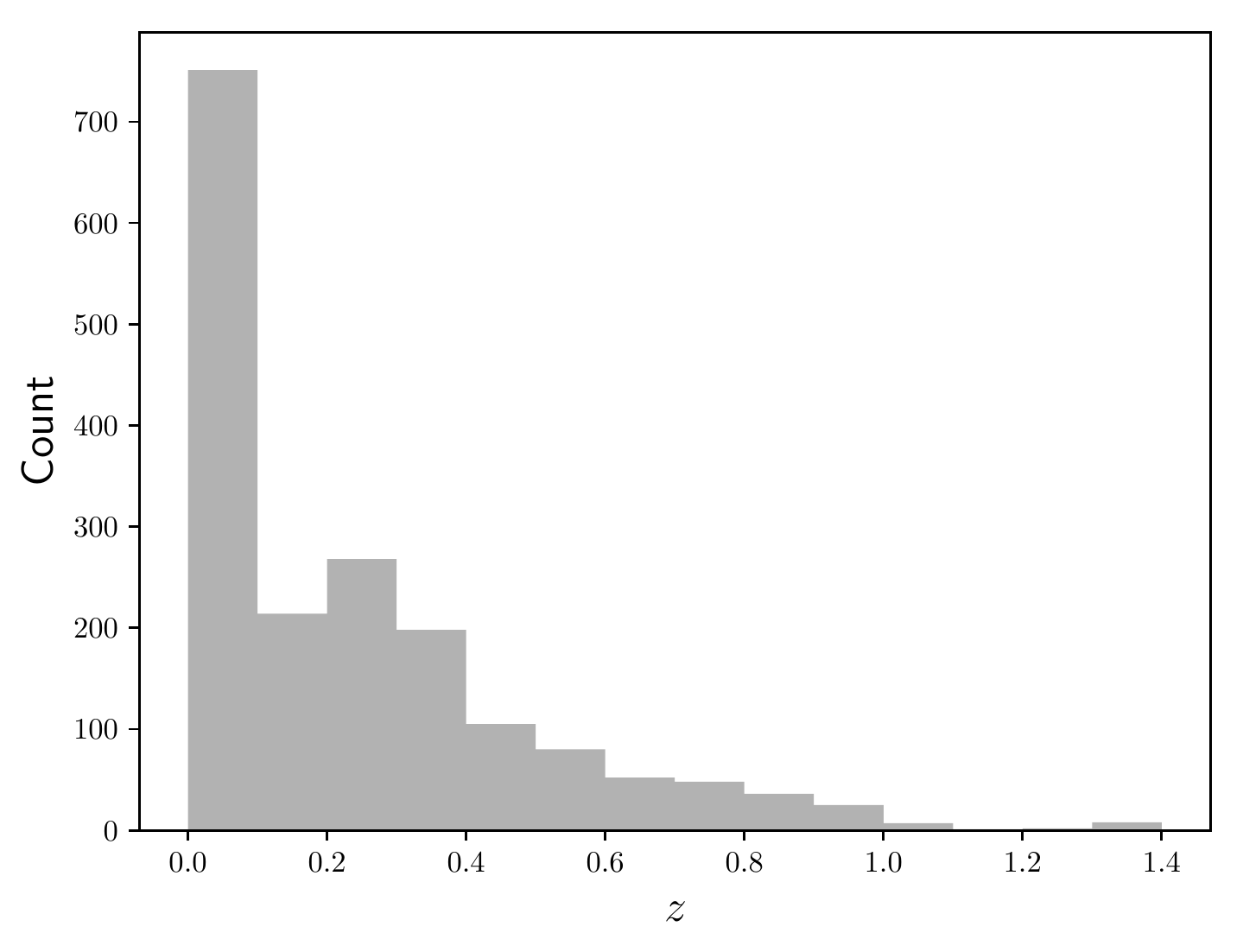}
\caption{The $z$ distribution of the combined Low-z, Foundation, SDSS, PS1, DES, and SNLS data sets.}
\label{fig:Redshift}
\end{figure}

The high-redshift sample is comprised of the Dark Energy 3-year sample \citep{Brout18a,Brout18b,Smith20}, the Sloan Digital Sky Survey (SDSS, \citealp{Sako18}), the Pan-STARRS survey (PS1MD \citealp{Rest14,Scolnic18}), the Supernova Legacy Survey (SNLS, \citealp{Betoule14}), and the Hubble Space Telescope (HST, \citealp{Riess2018}). Cross-calibration for all surveys is done in Brout et al. (\textit{in prep}). The redshifts for the entire Pantheon+ sample have been reevaluated in \cite{Carr21} and peculiar velocities are computed in \cite{Peterson21}.

In contrast to \cite{Brout21}, we require $z > 0.03$ to avoid difficulties modeling Hubble scatter with significant contributions from peculiar velocities. 

\subsection{Simulations}\label{sec:Simulations}

We use the SNANA simulation software \citep{SNANA, Kessler19} that broadly works in three steps: generation of fluxes from a source model, application of noise, and detection based on a characterisation of the difference imaging pipeline and/or spectroscopic selection. The simulation generates a rest-frame SED for each SNIa epoch and applies a combination of cosmological effects (such as dimming and redshift) and galactic effects (weak lensing, peculiar velocity, Galactic extinction). Next, the SED is integrated for each filter to determine broadband fluxes. Poisson noise is computed from the sky noise, source flux, PSF, and zero point. Next, the simulation models detection criteria and applies survey-dependent logic that requires a minimum number of detections separated in time; e.g., DES requires 2 detections found on different nights. Finally, a model for spectroscopic identification efficiency is applied.

We take our simulation inputs from the Pantheon+ analysis (Section 3 of \citealp{Brout21}). Table \ref{tab:SimInputTable} presents the source of instrumental inputs to generate each survey, which includes cadence and filter information (cadence library), single-visit detection efficiency vs. signal-to-noise ratio (DETEFF), and spectroscopic identification efficiency (SPECEFF).

\subsection{Selection}

Each sample has survey-specific selection requirements (cuts) as detailed in their data releases and Scolnic et al. (\textit{in prep}). Here we implement a more restrictive range of cuts to define the sample:
\begin{itemize}
    \item One observation at least five days before peak brightness in SN-rest frame, $T_{\rm rest} < 5 $
    \item Signal-to-Noise Ratio greater than 5 in two bands, $\textrm{SNR} > 5$
    \item Fitted stretch $-3 < x_1 < 3$
    \item Fitted colour $-0.3 < c < 0.3$
    \item Fitted $c$ uncertainty $c_{\rm err} < 0.1$
    \item Fitted $x_1$ uncertainty $x_{1_{\rm err}} < 1.5$
    \item Milky Way Reddening $E(B-V) < 0.3$ for low-redshift surveys 
    \item Fitted peak date uncertainty $\textrm{PKMJD}_{\rm err} < 20$ days\footnotetext[1]This is larger than conventional cuts of 2.0 days, but allows $\sim 10$ more SNIa into the fits.
    \item Light-curve fit probability $P_{\rm fit} >0.01$ for SDSS and $P_{\rm fit} > 0.001$ for DES and PS1. This $P_{\rm fit}$ cut is not applied to the other surveys$^1$. \footnotetext[2]{$P_{\rm fit}$ cuts are taken from the data releases of the specific surveys.}
    \item Chauvenet's criterion at $3.5 \sigma$ to distance modulus residuals relative to best fit cosmology
\end{itemize}

\section{Methodology}\label{sec:methodology}

\begin{figure*}[t]
\includegraphics[width=16cm]{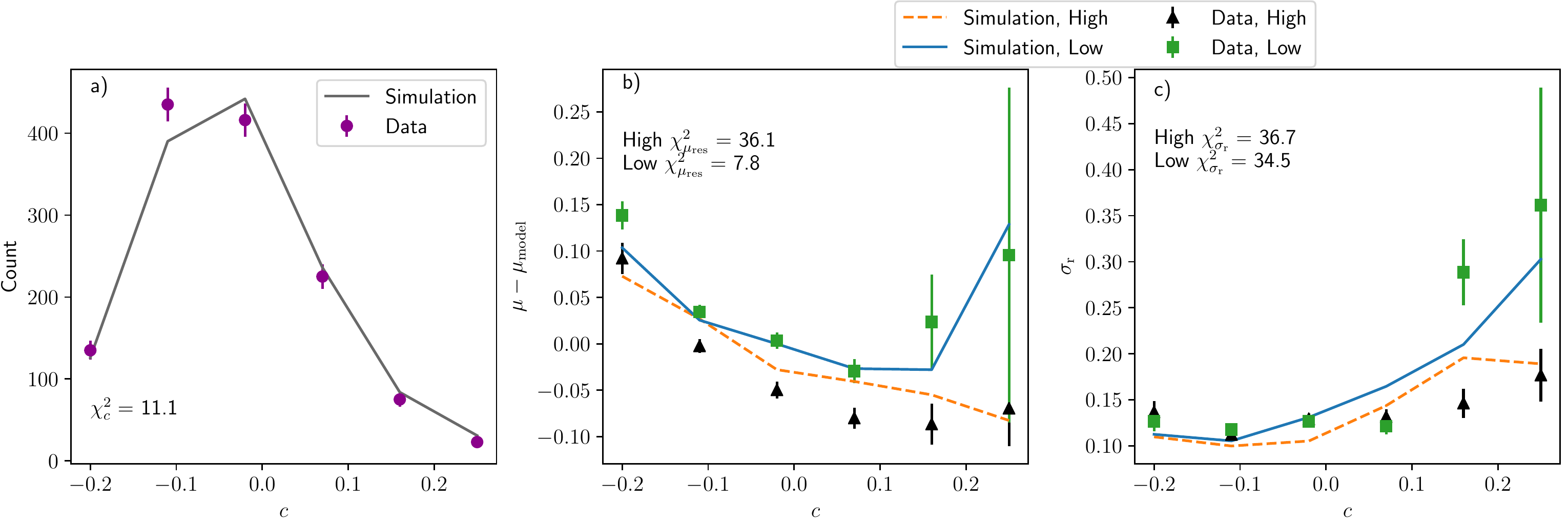}
\caption{Plots of the metrics described in Section \ref{sec:methodology:subsec:criteria}. The $c$ histogram, $c$ vs \mures, and $c$ vs \RMS\ are shown from left to right. The latter two plots are split on high and low mass. Plot a) is the observed colour ($c$) distribution with purple circles as histogram data and solid grey as simulated histogram. Plot b) is \mures\ vs $c$ split on high and low host-galaxy mass. Plot c) is \RMS~of \mures~vs. $c$. Green squares are low mass data and black triangles are high mass data, while blue plot is low mass simulation and orange dotted plot is high mass simulation. We can see good agreement between data and simulation for our best fit parameters.}
\label{fig:Criteria}
\end{figure*}

In contrast to \cite{Scolnic16} and \cite{Popovic21a}, which interpreted $c_{\rm obs}$ to come from a single underlying parameter, here the nature of the BS21 model is that several intrinsic populations (e.g $c_{\rm int}$, $R_V$, \EBV, and $\beta_{\rm SN}$) inform $c_{\rm obs}$. This added model complexity results in the need to forward model, where simulations generated from a set of intrinsic parameters are analysed in the same way as the data in order to compare the observed and simulated distributions. 

 \subsection{Metric Criteria}\label{sec:methodology:subsec:criteria}

We infer these dust-model parameters from observables that exist in conventional SNIa cosmological analyses. \mcmc\ uses the three constraints illustrated in Figure \ref{fig:Criteria}, evaluated in 6 colour bins ranging from $c = -0.2$ to $c = 0.25$ (encompassing more than 99\% of our sample). \cite{Scolnic16} used Figure \ref{fig:Criteria}a, the histogram of observed $c$ values; following BS21 we extend our constraints to include Hubble Residuals vs. $c$ (Fig. \ref{fig:Criteria}b) and Hubble Residual scatter vs. $c$ (Fig. \ref{fig:Criteria}c) to better constrain models with additional complexity. A single colour $\chi^2$ does not, by itself, contain enough information to separate the component pieces of the model. Broadly, the $\overline{R}_V$ value dictates the slope of the Hubble Residual values as a function of colour (Fig. \ref{fig:Criteria}b), and \RVstd\ drives both the colour-luminosity relation for the very reddest SNe (Fig. \ref{fig:Criteria}b), and separately, the Hubble Residual scatter vs. $c$ (\ref{fig:Criteria}c). The distribution of $\beta_{\rm SN}$ effects both the observed $\beta_{\rm SALT2}$, as well as providing a scatter floor for Hubble Residual scatter vs. $c$ (Fig. \ref{fig:Criteria}c). In addition to colour-dependent constraints, \mcmc\ includes a data-simulation constraint on $\beta_{\rm SALT2}$ and \sigint, though these do not have a graphical representation. The latter two constraints are determined by performing an `M11 Fit' \citep{Marriner11} at each MCMC step. For quantities that define the constraints, uncertainties are determined from the data, with the exception of \sigint, which is discussed in detail later in the section.

The first \mcmc\ constraint (Figure \ref{fig:Criteria}a) is a $\chi^2$ term of fitted $c$:

\begin{equation}\label{eq:chi2c}
\chi^2_c = \sum\limits_{i}(N^{\rm data}_{c_i} - N^{\rm sim}_{c_i})^2/e^2_{ni}
\end{equation}
where $N^{\rm data}_{c_i}$ is the number of SNIa in colour bin $i$ in the data, $N^{\rm sim}_{c_i}$ is the number of SNIa in colour bin $i$ in the simulation after scaling the integrated sum to match the data, and the uncertainty, $e_{ni}$, is 

\begin{equation}\label{eq:chi2cerr}
    e_{ni} = \sqrt{N^{\rm data}_{c_i}}
\end{equation}

The second constraint (Figure \ref{fig:Criteria}b) is based on the relationship between $c$ and the distance modulus residual \mures, 
\begin{equation}\label{eq:mures}
    \mu_{\rm res} = \mu - \mu_{\rm model}
\end{equation}
where $\mu$ is the measured distance modulus from Equation \ref{eq:tripp} and $\mu_{\rm model}$ is a reference cosmology. The Hubble residual constraint is

\begin{equation}\label{eq:chi2mures}
\chi^2_{\mu_{\rm res}} = \sum\limits_{i}(\mu_{\rm res_{i}}^{\rm data} - \mu_{\rm res_{i}}^{\rm sim})^2/e^2_{\mu_{\rm res_{i}}}
\end{equation}
where $\mu_{\rm \rm res_{ci}}$ is the \mures~in each $c$ bin and the error is 
\begin{equation}\label{eq:mureserr}
    e_{\mu_{\rm res_{i}}} = \frac{ {\sigma_{\rm r}}_{i} }{\sqrt{ {N_c}_{i} }}
\end{equation}
where ${\sigma_{\rm r}}_{i}$ is the robust scatter:
\begin{equation}\label{eq:robust}
    \sigma_{\rm r}(X) = 1.48 * \textrm{median}(|X|)
\end{equation}
and ${N_c}_{i}$ is the number of SNIa in each $c$ bin.

The third constraint (Figure \ref{fig:Criteria}c) is the relationship between $c$ and the Hubble residual scatter
\begin{equation}\label{eq:chi2RMS}
\chi^2_{\sigma_{\rm r}} = \sum\limits_{i}(\sigma_{\rm r_{i}}^{\rm data} - \sigma_{\rm r_{i}}^{\rm sim})^2/e^2_{\sigma_{\rm r} i}
\end{equation}
where $e_{\sigma_{\rm r} i}$ is defined as 
\begin{equation}\label{eq:robusterr}
    e_{\sigma_{\rm r}} = \frac{ {\sigma_{\rm r}}_{i} }{ \sqrt{{2N_c}_{i}} }
\end{equation}

from \cite{RPP} Equation 40.7.

\subsection{M11 Fit}\label{sec:methodology:subsec:absigma}

The colour-luminosity relationship, $\beta_{\rm SALT2}$, and the post-standardisation intrinsic scatter, \sigint, along with $\alpha_{\rm SALT2}$ and absolute luminosity $M(z_i)$, are determined from a global fit of the data following \cite{Marriner11} and Equation 3 in \cite{Kessler16}. This global fit computes $\alpha_{\rm SALT2}$, $\beta_{\rm SALT2}$, and \sigint\ by minimising Hubble scatter in bins of redshift. The M11 fit is done in reference to an arbitrary cosmological model in each $z$ bin, where the absolute rest-frame magnitude $M(z_i)$ is simultaneously determined in each $z$ bin. Thus we can assume that the reference $\mu_{\rm model}$ is accurate within each $z$ bin, but do not assume that the $\mu_{\rm model}$ is accurate across the entire redshift range. This M11 fit is performed at each step of the MCMC chain as part of the larger \mcmc\ method, and M11 fit results are used to evaluate the following \mcmc\ constraints:

\begin{equation}\label{eq:chi2beta}
\chi^2_{\beta_{\rm SALT2}} = (\beta_{\rm SALT2}^{\rm data} - \beta_{\rm SALT2}^{\rm sim})^2/e^2_{\beta_{\rm SALT2}}
\end{equation}
and
\begin{equation}\label{eq:chi2sigint}
\chi^2_{\sigma_{\rm int}} = (\sigma_{\rm int}^{\rm data} - \sigma_{\rm int}^{\rm sim})^2/e^2_{\sigma_{\rm int}}
\end{equation}
where $e^2_{\beta_{\rm SALT2}}$ is given by the M11 fit. M11 does not provide uncertainties for \sigint; therefore, we simulate 150 independent data-sized samples and estimate the uncertainty as the observed \sigint\ dispersion across these samples. We  find $e_{\sigma_{\rm int}} = 0.0036$.

The M11 fit is implemented by the Beams with Bias Corrections (BBC) code from \cite{Kessler16}, however, in our case, explicit bias corrections are not necessary because constraints (Eqs. \ref{eq:chi2c}, \ref{eq:chi2mures}, \ref{eq:chi2RMS}) are based on data and sims that are processed with the same M11 fit. Thus \mcmc\ accounts for selection effects from analysing simulated samples, not from BBC bias corrections.

\subsection{Dust2Dust Model Fitting}\label{sec:fastforward}

\begin{figure}[h]
\includegraphics[width=8cm]{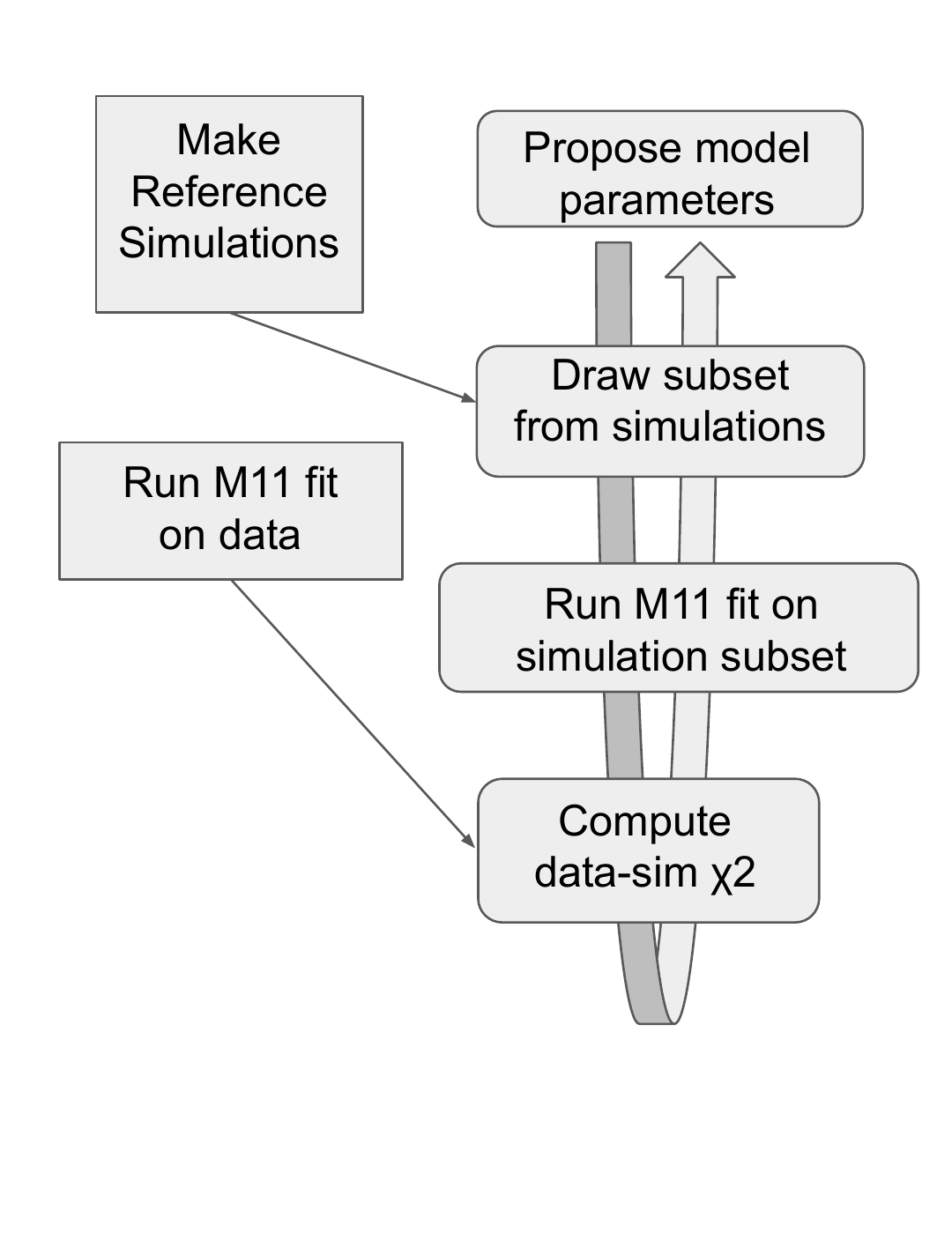}
\caption{Pictograph of the \mcmc~process. The requisite flat simulations are generated and SALT2mu \citep{Marriner11} is run on the data independently of the iterative process. The iterative steps are denoted with rounded corners; the model parameters from the BS21 model are proposed and a matching subset is drawn from the flat distributions. This subset then has the M11 run on it, and the results are compared to the data. }
\label{fig:Cartoon}
\end{figure}

Here, we improve on the BS21 fitting of parameters by developing a Markov Chain Monte Carlo (MCMC) program, named \mcmc, to both provide robust error modeling and the simultaneous fitting of the model parameters. The forward-modeling process contains two steps: An M11 fit is performed on the data and simulated supernova, and the results of this fit are used in the \mcmc\ likelihood to propose new steps in the MCMC chain.

A brief overview of the \mcmc\ process is provided in Figure \ref{fig:Cartoon} and summarised here:
\begin{itemize}
    \item Before running \mcmc, a reference simulation is created for \mcmc\ to quickly sample the parameter space.
    \item \mcmc\ proposes a specific set of model parameters from the BS21 model.
    \item \mcmc\ importance-samples from the reference simulation to create a weighted simulation characterised by the proposed set of model parameters.
    \item The M11 fit is run on both the data and \mcmc-created weighted simulation
    \item A data-sim $\chi^2$ is computed for the proposed model parameters and is used to inform the next proposed model parameters.
\end{itemize}

Details of the \mcmc\ process are as follows. \mcmc\ begins with generating reference simulations (as described in Section \ref{sec:Simulations}) containing bounding distributions of our model parameters, both SNIa and host-galaxy properties. Of note, the simulated $x_1$ distribution is excluded from the \mcmc\ process, as the BS21 model does not propose a relationship between $x_1$ and host-galaxy mass. Instead, the simulations are generated with stretch distributions taken from \cite{Popovic21a}, which include correlations with host-galaxy mass. These reference simulations are fitted with the SALT2 model and used in the MCMC process.

At each MCMC step, a new simulation is extracted by randomly selecting a subset of SNe from the reference simulation as follows. For SNIa parameters $\vec{\Theta} = \theta_i, i=1,2,3,4$, we define for each event
\begin{equation}
    \begin{split}
        p_{\rm ref} = \prod_{i=1}^{4} p_{\rm ref_{i}} \textrm{(Ref Sim Params)} \\
        p = \prod_{i=1}^{4} p_i \textrm{(MCMC Params)} ~~~
    \end{split}
\end{equation}
where $p_{\rm ref_{i}}$ is the bounding function probability for the $i_{\rm th}$ parameter used to generate the reference/bounding simulation, and $p_i$ is the probability of the SNe in the weighted simulation for the current MCMC step. The weight function for each event in the ref sim is

\begin{equation}
    P(X) = 
    \begin{cases} 
      ~ \textrm{if }\frac{p}{p_{\rm ref}} < U([0,1]), ~0& \\
      ~ \textrm{if }\frac{p}{p_{\rm ref}} \geq U([0,1]), ~1 
   \end{cases}
\label{eq:Reweight}
\end{equation}
where $U([0,1])$ is a random number in the range $[0,1]$. 

\begin{figure}[h]
\includegraphics[width=8cm]{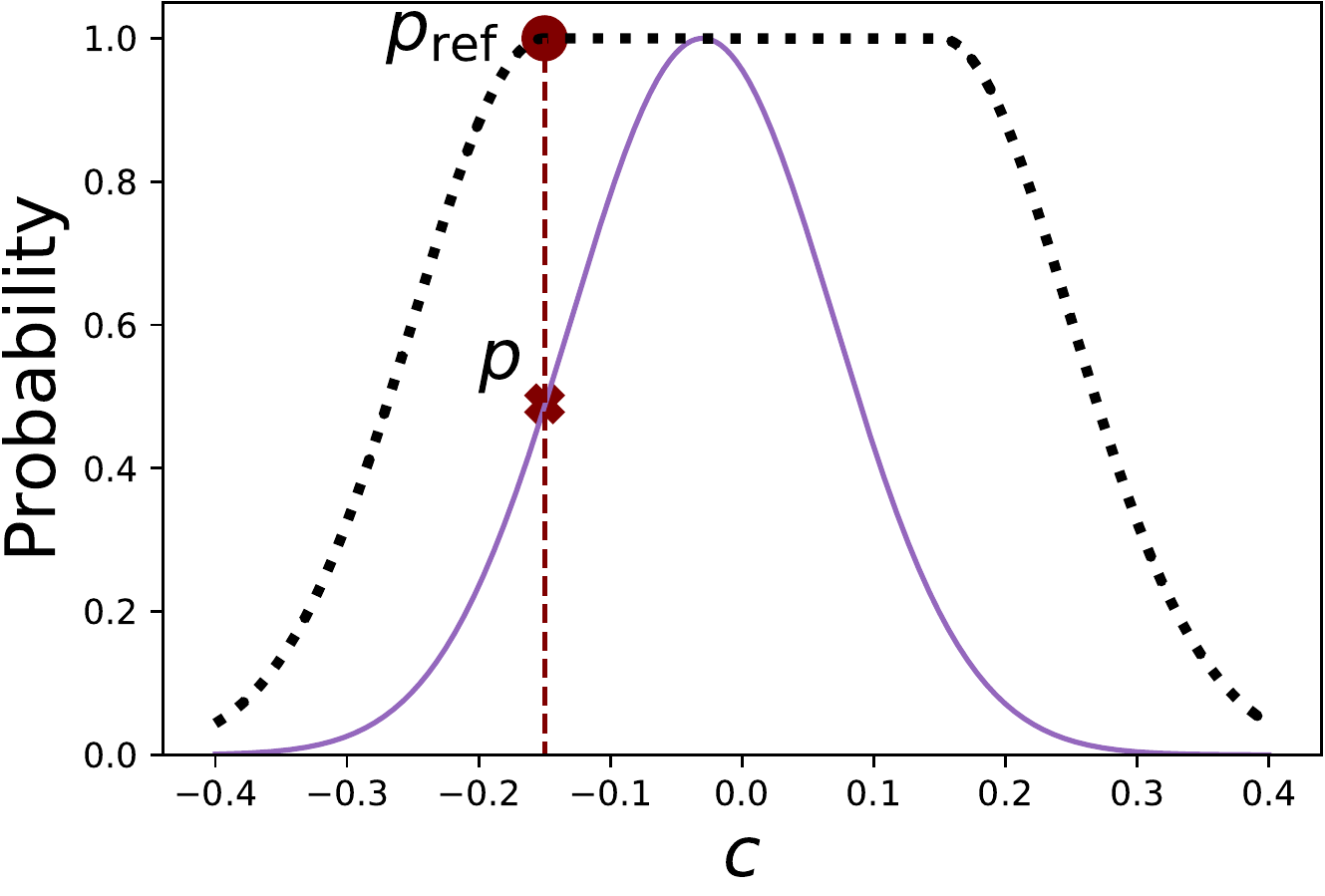}
\caption{An example of the reweighting done in \mcmc~for the $c$ parameter. The reference simulation is presented in black points and the proposed distribution is in purple histogram. The maroon $p_{\rm ref}$ point and $p$ X are compared, and simulated SNIa are kept according to Equation \ref{eq:Reweight}.}
\label{fig:Reweight}
\end{figure}

For \tautau, the bounding function is $(1/\tau)e^{-E_{\rm Dust}/\tau}$. For Gaussian distributions, our bounding function is chosen to be a `flat top' asymmetric Gaussian that is uniform between two boundaries, $\mu_1$ and $\mu_2$ (where $\mu_1 < \mu_2$), with standard deviations $\sigma_1$ and $\sigma_2$:
\begin{equation}
    P(X) = 
    \begin{cases} 
      ~ e^{-(X-\mu_1)^2/2\sigma^2_{1}} ~~ \textrm{if } X \leq \mu_1  & \\
      ~ 1 ~~~~~~~~~~~~~~~~~~~ \textrm{if} ~ \mu_1 < X < \mu_2 & \\
      ~ e^{-(X-\mu_2)^2/2\sigma^2_{2}} ~~ \textrm{if } X \geq \mu_2
   \end{cases}
\label{eq:Bounding}
\end{equation}
An example of this bounding function, along with the reweighting from Equation \ref{eq:Reweight}, is shown in Figure \ref{fig:Reweight}. Both the resulting distribution and the bounding function are normalised to have a peak probability of 1, though the bounding function does not integrate to unity by design.

This bounding function approach saves time by simulating fewer SNIa in parameter spaces that are unlikely to be used. The increase in efficiency can be approximated as the quotient of the areas covered by the bounding function and a uniform distribution, raised to the number of dimensions. In the case that all 4 of our reference distributions are bounded similarly to Figure \ref{fig:Reweight}, the quotient is $\sim 1.3$, and the computational efficiency is approximately $1.3^4 \sim 4$ times more efficient than uniform distributions. The combination of reference simulation, bounding functions, M11 fits, and the \texttt{emcee} package \citep{emcee} is the basis for our MCMC named \mcmc.

In the case that $p/p_{\rm ref} > 1$, the proposed distribution is considered to be out of range and flagged. The reference simulation can be iteratively regenerated with new centers and widths to cover the appropriate regions of parameter space. The number of such cases in our fit was on the order of $10^{-4}$, and therefore negligible. 

\mcmc\ initially keeps all simulated SNIa with a weight of 1. The resulting simulation size is heavily dependent on the proposed parameters - a tighter sigma beta cut will naturally have fewer supernovae than a wider one. In order to avoid statistical fluctuations, the simulation is downsized to a uniform selected size, in this case 5,000 supernovae. In the case that Dust2Dust is not able to downsize the sample, the results are flagged as being a bad fit and a log-likelihood of -infinity is returned. For example, with the given fiducial results of the paper, both $\sigma_{R_V}$ values can be as low as 0.3 without triggering the 5,000 SNe threshold. $\sigma_{\beta_{SN}}$ can be as low as 0.05 without triggering the 5,000 SNe threshold. The number 5,000 was chosen to minimise statistical fluctuations. Neither this downsizing nor the selection of model parameters significantly affect non-model distributions such as redshift or stretch.

The design of \mcmc\ is to use the simulation frameworks developed in \cite{SNANA, Kessler19, Kessler16}, and other works, while significantly reducing the simulation time compared to the naive approach of generating a new sample for each MCMC step. Minimising the time spent simulating is achieved via the use of the re-weighting algorithm, which generates an arbitrary desired distribution of SNIa properties after a single simulation. Table \ref{tab:Ranges} shows the boundaries for the walkers generated by \mcmc. These bounds were determined after a number of iterations using \mcmc\ to ensure that our choice of prior does not impact the results.

To fit our model, we minimise a modified version of Equation 8 in BS21:
\begin{equation}
        \begin{split} 
    \chi^2_{\rm Tot} = \chi^2_{\rm c} + \chilowrms +
    \chihirms  + \chilowres \\ 
    + \chihires + \chi^2_{\beta_{SALT2}} + \chi^2_{\sigma_{\rm int}}
   \end{split}
    \label{eq:chisqmcmc}
\end{equation}
where $\chi^2_{\rm c}$ is Equation \ref{eq:chi2c}, $\chilowrms$ and $\chihirms$ are Equation \ref{eq:chi2RMS} split on \mass, $\chihires$ and $\chilowres$ are Equation \ref{eq:chi2mures} split on \mass, and $\chi^2_{\beta_{SALT2}}$ and $\chi^2_{\sigma_{\rm int}}$ are Equations \ref{eq:chi2beta} and \ref{eq:chi2sigint} respectively. A summary of these criteria is presented in Table \ref{tab:LLExplanation}. 

\begin{table}[h]
\scalebox{1}{%
\begin{tabular}{c|c}
    $\chi^2_{\rm c}$ & $c_{\rm obs}$ distribution \\
     & Not split on \mass. \\
    \hline
     $\chihires$ & Colour vs. Hubble Residuals \\
      & For high-mass galaxies. \\
     \hline
     $\chilowres$ & Colour vs. Hubble Residuals \\
      & For low-mass galaxies. \\
    \hline
     $\chihirms$ & Colour vs. HR Scatter \\
      & For high-mass galaxies. \\
    \hline
     $\chilowrms$ & Colour vs. HR Scatter \\
      & For low-mass galaxies. \\
    \hline
     $\chi^2_{\beta_{SALT2}}$ & Error-weighted $\Delta$ $\beta_{SALT2}$ \\ 
      & Not split on \mass. \\
    \hline
     $\chi^2_{\sigma_{\rm int}}$ & Error-weighted $\Delta$ $\sigma_{\rm int}$ \\ 
      & Not split on \mass. \\ 
    \hline
\end{tabular}%
}
\caption{$\chi^2$ Terms in \mcmc\ Population Fit}
\label{tab:LLExplanation}
\end{table}

The updated simulation is compared to the data and a log-likelihood is computed to determine the goodness-of-fit. The log-likelihood is defined as
\begin{equation}\label{eq:LL}
    \mathcal{L} = -\frac{1}{2}*\chi^2_{\rm Tot} 
\end{equation}
which \texttt{emcee} maximises, equivalent to minimising $\chi^2_{\rm Tot}$.

For the purposes of this paper, we choose to split the BS21 model on host-galaxy \mass; however, any arbitrary external property can be used.

\begin{table}[h]
\scalebox{1}{%
\begin{tabular}{c|c|c}
    Parameter & Lower Bound & Upper Bound \\
    \cmean & $-0.3$ & $-0.04$ \\
    \cstd & $0.01$ & $0.2$ \\
    \RVmean \footnotetext[1]{The same ranges are used for low and high \mass}$^{a}$ & $0.8$ & $4$ \\
    \RVstd$^{a}$ & $0.25$ & $4$ \\
    \tautau\footnotetext[2]{The same ranges are used for low and high $z$}$^{ab}$ & $0.05$ & $0.2$ \\
    \betamean & $1.0$ & $3.0$ \\
    \betastd & $0.0$ & $1.0$ \\
    \hline
\end{tabular}%
}
\caption{Ranges for Each Model Parameter}
\label{tab:Ranges}
\end{table}

\section{Results} \label{sec:Results}

Here we evaluate and present the results from \mcmc\ run on the sample described in Section \ref{sec:data}. Section \ref{sec:Results:subsec:Fit} discusses the constraints on model parameters and how well they describe the data, including uncertainties and MCMC efficiency. Section \ref{sec:Results:subsec:w} reviews the impact of statistical and systematic model uncertainties on $w$.

\begin{figure*}[t]
\includegraphics[width=20cm]{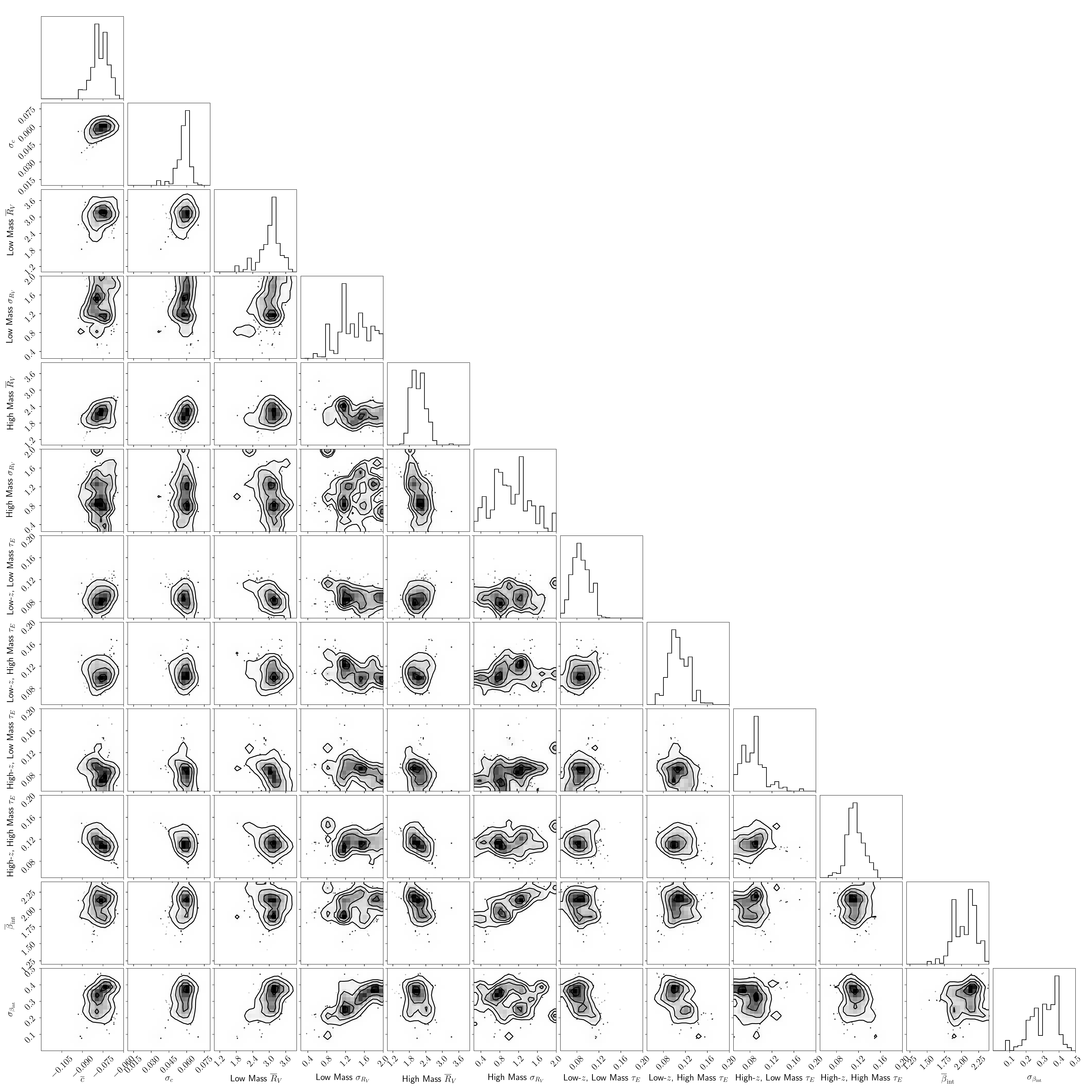}
\caption{Triangle plot for fitted \mcmc\ parameters. Parameters were fit simultaneously and include covariances. $\overline{c}$, $\sigma_c$, $\overline{\beta}$, and $\sigma_{\beta}$ are fit for the entire sample. $\overline{R}_V$, $\sigma_{R_V}$ are fit separately for low-mass and high-mass galaxies. $\tau_E$ is fit four times across $z$ and \mass. }
\label{fig:Corner}
\end{figure*}

\begin{table*}
\label{ptable}
\noindent
\caption{Results from Dust2Dust and Comparison With Original Model Parameters}

\makebox[.99\textwidth]{
\resizebox{.89\textwidth}{!}{
\begin{tabular}{|c|c|c|c|c|c|c|c|c|}
\hline\hline
Model & Sample &$\bar{c}$  & $\sigma_c$ & $\bar{\beta}_{\rm SN}$ & $\sigma_{\beta_{\rm SN}}$  & $\bar{R}_V$ & $\sigma_{R_V}$ & $\tau_E$  \\
\hline

\textbf{Mean:}&&&&&&&&\\
High-mass & CfA, CSP, Foundation&
\MELHcm&
\MELHcs&
\MELHbm&
\MELHbs&
\MELHRm&
\MELHRs&
\MELHTe
\\

High-mass & DES, PS1, SNLS, SDSS   &
 &
 &
 &
 &
 &
 &
\MEHHTe
\\
Low-mass & CfA, CSP, Foundation&
 &
 &
 &
 &
\MELLRm&
\MELLRs&
\MELLTe
\\

Low-mass & DES, PS1, SNLS, SDSS   &
 &
 &
 &
 &
 &
 &
\MEHLTe
\\
\hline

\textbf{Original:}&&&&&&&&\\

High-mass & CfA, CSP, Foundation&
$-0.084 \pm 0.004$&
$0.042 \pm 0.002$&
$1.98 \pm 0.180$&
$0.35 \pm 0.200$&
$1.50 \pm 0.250$&
$1.300 \pm 0.200$&
$0.190 \pm 0.080$
\\

High-mass & DES, PS1, SNLS, SDSS   &
 &
 &
 &
 &
 &
 &
$0.150 \pm 0.020$
\\
Low-mass & CfA, CSP, Foundation&
 &
 &
 &
 &
$2.75 \pm 0.350$&
$1.300 \pm 0.200$ &
$0.010 \substack{+0.050 \\ -0.010}$
\\

Low-mass & DES, PS1, SNLS, SDSS   &
 &
 &
 &
 &
 &
 &
$0.120 \pm 0.020$
\\
\hline

\end{tabular}%
}}

\vspace{.03in}

\label{tab:paramtable}
\end{table*}

\begin{table}[h]
\scalebox{1}{%
\begin{tabular}{c|c|c|c|c}
    $\chi^2$ & G10 & C11 & BS21 & \mcmc\ \\
     & & & Original & \\
    \hline
    $\chi^2_{\rm c}$        & 37  & 35  & 68  & $11$ \\
    $\chihires$        & 58  & 54  & 18 & $36$  \\
    $\chilowres$         & 19  & 54  & 20 & $8$ \\
    $\chihirms$          & 158 & 162 & 95 & $37$ \\
    $\chilowrms$            & 189 & 157 & 56 & $35$ \\
    $\chi^2_{\beta_{SALT2}}$ & 7   & 31  & 1 & $0$ \\
    $\chi^2_{\sigma_{\rm int}}$                & 4   & 25  & 27 & $3$ \\
     & & & & \\
    Sum & 472 & 518 & 285 & 129 \\
\end{tabular}%
}
\caption{Breakdown of different $\chi^2$ criteria for the G10, C11, BS21-Original, and \mcmc\ scatter models.}
\label{tab:LLTable}
\end{table}

\subsection{Fit of Model Parameters} \label{sec:Results:subsec:Fit}

The results for our best-fit parameters are given in Table \ref{tab:paramtable}, and these results are discussed below. Figure \ref{fig:Corner} shows the posterior likelihoods for our model parameters after discarding \burnin ~steps for burn-in. After this burn-in, our chains comprised 18,900 steps with 24 walkers. The acceptance fraction is 15\%, and the autocorrelation time is 100. While this autocorrelation time could be improved by running \mcmc\ longer, we are already near the computation limits of our setup. In general, the posteriors are uni-modal, though not well approximated by a Gaussian. We give the mean values from the chains as our fiducial result.

\begin{figure*}
\textbf{G10 Scatter Model}\par\medskip
\includegraphics[width=16cm]{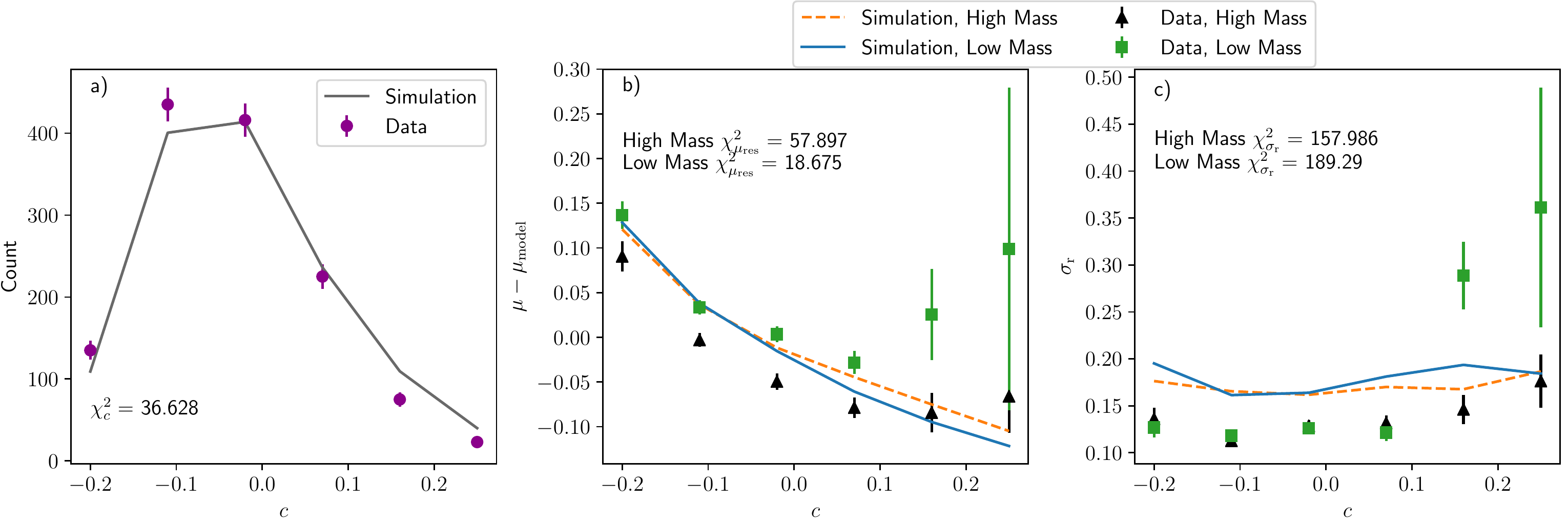}
\caption{The $c$ histogram, $c$ vs \mures, and $c$ vs RMS(\mures) are shown here from left to right. The latter two plots are split on high and low mass. Purple circles are histogram data and solid grey is simulated histogram. Green squares are low mass data and black triangles are high mass data, while blue plot is low mass simulations and orange dotted plot is high mass simulation. The G10 model significantly overestimates the RMS and does not accurately replicate the \mures~split nor the change in RMS with $c$. }
\label{fig:G10-Criteria}
\end{figure*}

\begin{figure*}
\textbf{C11 Scatter Model}\par\medskip
\includegraphics[width=16cm]{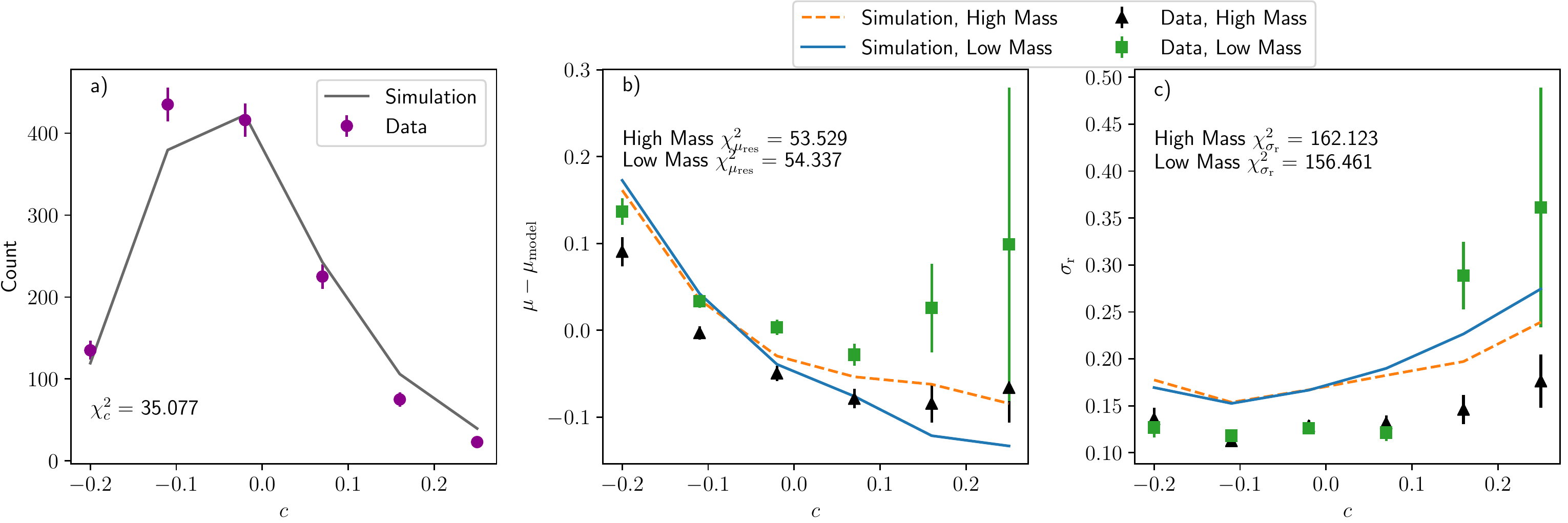}
\caption{The $c$ histogram, $c$ vs \mures, and $c$ vs RMS(\mures) are shown here from left to right. The latter two plots are split on high and low mass. Purple circles are histogram data and solid grey is simulated histogram. Green squares are low mass data and black triangles are high mass data, while blue plot is low mass simulations and orange dotted plot is high mass simulation. The C11 model significantly overestimates the RMS and does not accurately replicate the \mures~split.  }
\label{fig:C11-Criteria}
\end{figure*}

As described in Section \ref{sec:methodology:subsec:criteria}, Figure \ref{fig:Criteria} shows three of the five criteria used to constrain the BS21 model parameters. A more specific breakdown of the $\chi^2$ values is shown in Table \ref{tab:LLTable}. \mcmc\ returns a $\chi^2$ that is $\times 2$ smaller than the original BS21 (BS21-Original) and nearly $\times 5$ smaller than two commonly used scatter models, \citealp{Guy10} (G10) and \citealp{Chotard11} (C11). Both G10 and C11 follow SALT2 in not including an explicit dust-based colour law; G10 ascribes approximately 75\% of observed scatter to achromatic effects and the remaining 25\% to chromatic effects. C11, in contrast, attributes 75\% of scatter to chromatic variations and the remaining 25\% grey. The equivalent criteria of Figure \ref{fig:Criteria} for the G10 and C11 scatter models are shown in Figures \ref{fig:G10-Criteria} and \ref{fig:C11-Criteria}, respectively. For both the G10 and C11 models, the Huubble Residual and $\sigma_{\rm r}$ distributions are poorly recreated by simulations.

\mcmc\ finds model parameters such that simulations predict the observed $c$ distribution and \mures\ behaviour, as well as $\beta_{\rm SALT2}$ and $\sigma_{\rm int}$. The Hubble residual RMS term ($\chi^2_{{\sigma}_r}$) is the largest contributor to the overall $\chi^2$.  

We test the robustness of \mcmc\ by analysing 16 data-sized simulations using parameters taken from the means in Table \ref{tab:paramtable} and compare the aggregate results. The resulting $\chi^2_{\rm tot}$ ranges from 42 to 105, with an average of 70. These $\chi^2_{\rm tot}$ are significantly smaller than for the data ($\chi^2_{\rm tot} = 129$), which suggests that improvements to the model, notably $\sigma_{\rm r}$, are likely needed.

The 12 fitted parameters are consistent with simulated inputs to within their uncertainties. Similar to our fiducial result, we use the mean values from the chains as the results for the 16 fits to data-sized simulations. From these fits, we compute the standard deviation of parameter values to compare to our estimated uncertainties. We find the medians of the 16 posterior widths agree to within 15\% of the scatter ($\sigma$) of the 16 posterior means: the errors reported for the parameter fits are accurate to within 15\% of the standard deviation of the 16 best fits, though they are underestimated.  We then check how consistent the posterior means are to the true value of each parameter.  We find of the 12 parameters, only one shows a $2.9\sigma$ bias - this is the low mass $\sigma_{Rv}$ parameter. The median value of the distribution of recovered low mass $\sigma_{Rv}$ values is within 2$\sigma$ of the input value when using the medians of the posterior widths. To verify this, we ran on a single large simulated sample of 50,000 SNe to check for parameter recovery biases. Three of our recovered parameters are more than $1\sigma$ from the input values; none are greater than $2\sigma$. 

Under the assumption of our model, we find evidence of different $R_V$ distributions as a function of host-galaxy mass. The mean $\overline{R}_V$ values in Table \ref{tab:paramtable} differ between high and low mass galaxies by $2.8\sigma$, and the \RVstd\ values are within $1\sigma$.

We find evidence that the $\beta_{\rm SN}$ distribution is not a delta function: $\sigma_{\beta_{\rm SN}} > 0$ with $>3 \sigma$ confidence for the mean \mcmc\ results. G10 and C11 both assume a delta function for $\beta_{\rm SALT2}$, which is not consistent with our \mcmc\ results.

\subsection{Comparison with Original BS21 Parameters}\label{sec:Results:subsec:Comparison}

For \cmean\ (Table \ref{tab:paramtable}), we find peak probability (\MELHcm) comparable to that of BS21-Original ($0.084 \pm 0.004$), though the standard deviation (\MELHcs) is higher than that of BS21-Original: $0.042 \pm 0.002$.

Our \betamean\ and \betastd\ values are consistent with BS21-Original, and our values are more precise. Our \betastd\ uncertainty is $\sim 4\times$ smaller. Our \betamean~is \MELHbm, compared to $1.98 \pm 0.18$. 

In contrast to BS21-Original, who found a low-mass \RVmean\ value at $1.5 \pm 0.25$, we find low-mass \RVmean\ to be $\sim 2.0 \pm 0.25$. Our high-mass \RVmean\ ($\sim 3.0 \pm 0.4$) is consistent with BS21-Original.

The \mcmc-recovered \RVstd\ results are lower or consistent with those in BS21-Original. While the low-mass \RVstd\ are consistent, our high-mass \RVstd\ is smaller compared to BS21-Original. 

\cite{BS20} found that \tautau\ has little impact on the overall \logl, particularly for the low-$z$ subsample (BS21 Figure 9). Using \mcmc\ and a larger data sample, we find that all four \tautau\ values are similar (0.08 - 0.11), but the low and high mass components are more self-similar than the original BS21.

\subsection{Systematic Uncertainties on Cosmological Parameters} \label{sec:Results:subsec:w}

For the SNIa-cosmology analysis (Brout et al. \textit{in prep.}, B22), the fitted \mcmc\ parameters are used in simulations for bias corrections, and here we evaluate the associated systematic uncertainty on the dark energy equation-of-state $w$, matter density $\Omega_M$, and the Hubble Constant H$_0$. We compare two systematic evaluation methods: 1) a forward modeling approach using 100s of simulations, each analysed with a fast cosmology $\chi^2$ minimiser, and 2) a covariance matrix approach with wfit.

We perform a cosmology analysis that includes 
\begin{itemize}
    \item SALT2 light-curve fitting to standardise the brightness (Section \ref{sec:Model:subsec:SALT})
    \item Generating a large simulated sample for bias corrections to account for selection effects
    \item using Beams with Bias Corrections (BBC, \citealp{Kessler16, Popovic21a}) to produce a Hubble diagram corrected for biases in 4 dimensions: $\{c, z, x_1, M_{\rm stellar}\}$
    \item A cosmology fit
\end{itemize}

The first three steps are performed in the same manner as in B22. For our blinded cosmology fitting, we use a simple and fast minimisation program in SNANA (wfit) that uses a Planck-like Cosmic microwave background (CMB) prior based on the R-shift parameter (see $\sigma_{\rm R}$ tuning discussion at the end of Section 3 in \citealp{Sanchez21}).

To model the systematic uncertainty in $w$ from BS21 parameter uncertainties, we generate \numsim\ statistically independent data-sized simulations with BS21 model parameters drawn randomly from \mcmc\ chains, thus accounting for the uncertainty and covariance between model parameters. To account for the statistical uncertainty, we additionally generated \numsim\ statistically independent data-sized simulations with best fit BS21 model parameters. Each data-sized simulation is analysed using the 4 steps above.

\newcommand{\wsysnstat}{\textrm{STDEV}(w_{\rm syst + stat})}
\newcommand{\wsys}{\textrm{STDEV}(w _{\rm stat})}
\newcommand{\wsig}{\sigma w_{\rm sys}}

We define the systematic uncertainty from population parameters to be
\begin{equation}\label{eq:wsys}
   \wsig = \sqrt{\wsysnstat^2-\wsys^2}
\end{equation}
where $\wsysnstat$ is the standard deviation of $w$ values from \numsim\ simulations including the randomly drawn parameters, and $\wsys$ is the standard deviation of $w$ values from \numsim\ simulations using the best fit parameters. We find $\wsig = 0.014 \pm 0.001$.

While this simulation approach is effective for evaluating a single systematic, it is impractical for the commonly-used covariance matrix method that is used in the Pantheon+ cosmology analysis. Instead, we attempt to model the systematic uncertainty in the covariance matrix method using 4 bias correction perturbations that mimic the \numsim\ simulations approach.

For this covariance matrix approach, we generate the nominal bias correction simulation using the maximum likelihood set of parameters. Next, we generate three bias correction simulations with model parameters that are $\Delta \chi^2 \sim 10$ (approximately $1\sigma$ away from the mean) away from our best-fit results. The resulting Hubble diagram and  statistical$+$systematic covariance matrix is processed through wfit to determine cosmological parameters using a CMB prior from \cite{Planck16}. A summary of this process is given in Section 2.2 of B22. While our forward modeling method only evaluated the $w$ systematic, here we evaluate the systematic uncertainty on both $w$ and H$_0$. 

\newcommand{\whsysnstat}{\sigma w_{\rm syst+stat}}
\newcommand{\whsys}{\sigma w _{\rm stat}}
\newcommand{\hsigma}{\sigma \textrm{H}_{0\rm sys}}
\newcommand{\hsysnstat}{\sigma(\textrm{H}_{0\rm syst + stat})}
\newcommand{\hsys}{\sigma(\textrm{H}_{0\rm stat})}
\newcommand{\hunit}{km/s/Mpc}

We define the systematic uncertainties from population parameters to be
\begin{eqnarray}\label{eq:twosys}
   \wsig &=& \sqrt{\whsysnstat^2-\whsys^2} \\
    \hsigma &=& \sqrt{\hsysnstat^2-\hsys^2}
\end{eqnarray}
where $\whsysnstat$ is the $w$-uncertainty from wfit using a systematic+statistic covariance matrix, and $\whsys$ is the statistical-only $w$-uncertainty. Analogous definitions apply to $\hsysnstat$ and  $\hsys$. We find $\wsig = \covbinw$, and $\hsigma = 0.145$~\hunit. 

The results from the forward-modeling and covariance matrix methods, while not within error, are similar in their impact on $w$. With only three different biasCor simulations, our covariance matrix approach may be affected by noise. Adding more biasCor simulations represents a significant computational challenge. To maintain consistency with the forward-modeling approach, we binned the Hubble Residuals for the covariance matrix approach. Implementing an unbinned approach from \cite{Binning} allows the data to ``self correct" for certain systematics, and may decrease this systematic uncertainty fro m population modeling.
The forward-modeling is therefore an upper bound on the associated systematic uncertainty.

\begin{table}
    \centering
    \caption{Cosmology constraints with systematics with $\Omega_{\rm M}$ prior}\label{tab:H0OMw}
    \scalebox{1}{%
    \begin{tabular}{ccc}
        \hline
		Model & $\sigma$H$_0$ & $\sigma w_{\rm sys}$\\
		\hline
		Forward & - &  0.014(1) \\
        Covariance & 0.145~\hunit & \covbinw \\
		\hline
    \end{tabular}
    }
\end{table}

\section{Discussion, Future Work, and Conclusion}

Although we have used \mcmc ~ to fit model parameters based on \mass, any external host property can, in principle, be used. Future works can fit for parameters based on other properties such as galaxy morphology or sSFR. Meldorf et al. 2021 (\textit{in prep}) shows that fitting on sSFR may prove promising for determining total extinction and reddening in SNIa samples. 

Our SNIa model assumes a single intrinsic colour-luminosity relationship ($\beta_{\rm SN}$) and differing $R_V$ distributions ($\Delta \overline{R}_V = 1$) to explain the mass step. \cite{Gonzalez-Gaitan21} take an alternative approach, assuming no dust contribution an instead investigating differing $\beta_{\rm SN}$, though they do not find this assumption sufficient to explain the mass step. While we find evidence of differing $R_V$ distributions across low and high mass galaxies $(2.8\sigma)$, studies such as \cite{Thorp21} find no significant $R_V$ difference. We have chosen to assume a single intrinsic colour distribution for all SNIa with differing underlying $E(B-V)$ distributions as a function of host properties, in line with the theory provided by BS21. However, future analyses using \mcmc\ may wish to further investigate this claim by allowing the intrinsic colour to vary with host galaxy \mass.

With regards to the mass step, \cite{Johansson20} find that a $\Delta \overline{R}_V = 0.5$ difference is sufficient to explain the mass step in optical bands, and furthermore, do not find a mass step in the NIR. It is worth noting that \cite{Johansson20} were not able to replicate the observed NIR mass step presented in \cite{Uddin20}, nor that of \cite{Ponder20}, though  NIR data analysed by \cite{Jones22} observe consistent findings to \cite{Ponder20} and \cite{Uddin20}.

Our forward-modeling approach was chosen over other potential Bayesian frameworks due to difficulties in modeling systematic errors that inhere in traditional Bayesian approaches. The choice to use simulations allows us to account for selection effects and parameter covariances that are difficult to analytically model in a Bayesian approach. Preserving these selection effects and covariances makes for more accurate parameter modeling and the ability to estimate systematic uncertainties.

We find a set of valid BS21 model parameters and that our choice in model parameters has a small ($w_{\rm sys} < 0.01$) impact on $w$ and a negligible impact on H$_0$. This $w$-impact is smaller than previous estimates for intrinsic populations and choice of scatter model \citep{Brout18SYS}. While continued work is needed to further constrain $R_V$ and \EBV\ distributions, these results are promising for further research into the effects of host-galaxy properties on SNIa samples, and other avenues of research could further inform our priors.

Future photometric samples, while not statistically limited with the implementation presented in this approach, present issues for more complex modeling of parameters and subsamples. The most notable of these issues is the potential of core-collapse supernovae contamination in the Ia sample, biasing recovered model parameters; concerns with properly identifying host galaxies and redshifts will also present problems. We expect the increased statistics for future surveys such as LSST-SN and the Nancy Grace Roman Supernova Survey to allow improved models to account for changing parameters across the chosen external parameter. Nonetheless, the change from spectroscopic to photometric identification will present its own issues.

\mcmc\ will also be a crucial tool in future cosmological analyses. Recalibration and the retraining of light-curve fitting tools, such as SALT2 \citep{Taylor21} and SALT3 \citep{Kenworthy21}, can significantly affect scatter model parameters. This shift in calibration and light-curve fitting will necessitate inferring new scatter model parameters for use in generating bias correcting simulations. As such, these new surveys will require implementing \mcmc\ into their cosmological pipelines to significantly reduce unnecessary overhead.

\section{Acknowledgments}

BP acknowledges the contributions of Courtney Hazen and Helen Qu in the development of \mcmc; along with GT for general discussions. DB acknowledges support for this work was provided by NASA through the NASA Hubble Fellowship grant HST-HF2-51430.001 awarded by the Space Telescope Science Institute, which is operated by Association of Universities for Research in Astronomy, Inc., for NASA, under contract NAS5-26555. DS is supported by DOE grant DE-SC0010007, DE-SC0021962 and the David and Lucile Packard Foundation. DS is supported in part by the National Aeronautics and Space Administration (NASA) under Contract No.~NNG17PX03C issued through the Roman Science Investigation Teams Program.  This work was completed in part with resources provided by the University of Chicago’s Research Computing Center. Simulations, light-curve fitting, BBC, and cosmology pipeline managed by \texttt{PIPPIN} \citep{PIPPIN}. Analysis and visualisations provided in part by https://github.com/bap37/Midwayplotter. 

\newpage

\bibliographystyle{mne2.bst}

\bibliography{research2.bib}

\end{document}